\documentstyle[12pt]{article}
\newcommand{\Ref}[1]{{\rm(\ref{#1})}}
\newcommand{\C}{{\bf C}}

\newcommand{\R}{{\bf R}}
\newcommand{\RR}[2]{R(\lambda_{#1#2})^{(#1#2)}}
\newcommand{\Z}{{\bf Z}}
\newcommand{\ben}{\begin{equation}}
\newcommand{\een}{\end{equation}}
\newcommand{\bean}{\begin{eqnarray}}
\newcommand{\eean}{\end{eqnarray}}
\newcommand{\be}{\begin{displaymath}}
\newcommand{\ee}{\end{displaymath}}
\newcommand{\bea}{\begin{eqnarray*}}
\newcommand{\eea}{\end{eqnarray*}}
\newcommand{\Root}[3]{\put(0,0){\line(#1,#2){#3}}}

\newcommand{\sn}{{\rm sn}}

\newenvironment{proof}{\noindent{\em Proof\/}:}{$\;\Box$}

\newenvironment{example}{\par\vspace{.5\baselineskip}
\noindent{\em Example\/}:}{}
\newtheorem%
{thm}{Theorem}
\newtheorem%
{proposition}[thm]{Proposition}
\newtheorem%
{lemma}[thm]{Lemma}
\newtheorem%
{corollary}[thm]{Corollary}

\title{Elliptic Dunkl operators, root systems, and functional equations}
\author{V. M. Buchstaber${}^*$, G. Felder${}^{**}$ and A. P. Veselov${}^*$}
\date{March 1994}
\begin{document}
\maketitle
\centerline{${}^{*}$Department of Mathematics and
Mechanics, Moscow State University,}
\centerline{119899 Moscow, Russia}
\smallskip
\centerline{${}^{**}$Department of Mathematics, University of North Carolina
at Chapel Hill,}
\centerline{Chapel Hill, NC 27599-3250, USA}
\smallskip

\subsubsection*{Introduction}
In the paper \cite{Du}, Dunkl introduced the following
difference-differential operators acting on functions
on a Euclidean space $V$, related to arbitrary
finite groups $G$ generated by orthogonal reflections in
$V$:
\ben\label{e1}
\nabla_\xi=\partial_\xi+\sum_{\alpha\in R_+}
k_\alpha(\alpha,\xi){1\over (\alpha,x)}\hat s_\alpha.
\een
Here $\partial_\xi$ denotes the partial
derivative in direction $\xi\in V$,
$R$ is the root system of the group $G$, i.e., the set
of unit normals to the reflection hyperplanes, $R_+$ is
its positive part with respect to some generic linear form on $V$,
$k_\alpha=k(\alpha)$ is a $G$-invariant function on $R$,
$s_\alpha$ is the reflection corresponding to the root $\alpha\in R$,
and $\hat s_\alpha$ is the operator on the space of functions on
$V$:
\be
\hat s_\alpha f(x)=f(s_\alpha(x)).
\ee
To be precise, Dunkl used slightly different
operators, which are conjugated to \Ref{e1}
by the operator of multiplication by $\prod(\alpha,x)^{k_\alpha}$.

The main property of the Dunkl operators
is given by the following

\begin{thm} (Dunkl)
The operators \Ref{e1} commute with each other:
\ben\label{e2}
[\nabla_\xi,\nabla_\eta]=0,
\een
for all $\xi$, $\eta\in V$.
\end{thm}

The goal of this work is to describe certain generalizations
of the Dunkl operators \Ref{e1}, preserving the property
\Ref{e2}. Some of these results were announced in \cite{Ve}.

 In Section 1 we consider generalizations of the form
\ben\label{e3}
\nabla_\xi=\partial_\xi+\sum_{\alpha\in A_+}
k_\alpha(\alpha,\xi){1\over (\alpha,x)}\hat s_\alpha,
\een
where $A_+$ is the set of unit normals to some set $S$
of hyperplanes in $V$ passing through the origin,
$A_+$ is its positive part, and $k_\alpha=k(\alpha)$ is
some function on $A_+$. We show that the commutativity
of the operators $\nabla_\xi$ implies that $S$ is the set
of reflection hyperplanes of some Coxeter group $G$,
$A_+=R_+$, and $k$ is  $G$-invariant.

 In the Section 2, we consider operators of the form
\ben\label{e4}
\nabla_\xi=\partial_\xi+\sum_{\alpha\in R_+}
(\alpha,\xi)f_\alpha((\alpha,x))\hat s_\alpha,
\een
where $f_\alpha(z)$ are functions of one variable,
not identically $0$.

The commutation relations \Ref{e2} are equivalent
to a system of functional equations for the functions
$f_\alpha$, $\alpha\in R_+$.
In the case when $G$ is the Weyl group $W$ of a simple Lie
algebra, we show that, with the exception of
$A_1$, $B_2$, the only $W$-invariant solutions,
i.e., such that
\be
 \hat s_\alpha\nabla_\xi=\nabla_{s_\alpha(\xi)}\hat s_\alpha,
\ee
for all $\alpha\in R_+$, are Dunkl's solutions, in accordance
with \cite{Ch}.

While the $A_1$ case is trivial, the $B_2$ case
leads to the classical theory of  {\em Landen's and Jacobi's
transformations} of elliptic functions. We give the general
$W$-invariant solution in this case. This result
has  an interesting topological application in the theory
of elliptic genera (see \cite{BuVe}).

In Section 3, we give a solution of
the functional equations in terms of elliptic functions,
for arbitrary reduced root systems. This solution gives
families of commuting differential-difference operators
that we call elliptic Dunkl operators.

We then show in Section 4, using techniques from \cite{Bu}, \cite{Bu2},
that these are essentially all solutions in the
$A_{n-1}$ case.

In Section 5, quantum elliptic Dunkl operators are introduced.
These are pairwise commutative families
of difference operators. They depend on a parameter $\mu$, and
are such that elliptic Dunkl operators appear in the
first order term (semiclassical approximation) of their
expansion in power of $\mu$. These operators
are related to the transfer matrices associated with
the  $R$-matrix of \cite{ShiUe}, \cite{FePa}.

We conclude by discussing the possible applications
of our results to the theory of integrable $n$-body
systems.

\subsubsection*{1. Operators of Dunkl type and Coxeter groups}

Let $S$ be a finite set of hyperplanes in a Euclidean
space $V$ passing through the origin, $A$ the set of
unit normals to the hyperplanes in $S$
 (two for each hyperplane), $A_+$ the positive half
of $A$ with respect to some linear form on $V$, and
$k_\alpha$ some non-zero coefficients.

Let $\nabla_\xi$ be the operator
\ben\label{e5}
\nabla_\xi=\partial_\xi+\sum_{\alpha\in A_+}
k_\alpha(\alpha,\xi){1\over (\alpha,x)}\hat s_\alpha,
\qquad k_\alpha\neq 0.
\een

\begin{thm}
The operator $\nabla_\xi$ and $\nabla_\eta$ commute
for arbitrary $\xi$ and $\eta\in V$ if and only if
$A_+$ coincides with $R_+$ for some Coxeter group
$G$ and $k(\alpha)=k_\alpha$ is a $G$-invariant
function.
\end{thm}

\begin{proof}
The commutator $[\nabla_\xi,\nabla_\eta]$ can be
rewritten in the form I+II+III (see \cite{He}),
where
\bea
{\rm I}&=&[\partial_\xi,\partial_\eta]=0,\\
{\rm II}&=&\partial_\xi(\sum_\alpha k_\alpha
\frac{(\alpha,\eta)}{(\alpha,x)}\hat s_\alpha)
-
\partial_\eta(\sum_\alpha k_\alpha
\frac{(\alpha,\xi)}{(\alpha,x)}\hat s_\alpha)=0,
\,\\
{\rm III}&=&
\sum_{\alpha,\beta\in A_+}
k_\alpha k_\beta
\langle \alpha,\beta\rangle_{\xi,\eta}
\frac1{(\alpha,x)(s_\alpha(\beta),x)}\hat s_\alpha\hat s_\beta,
\eea
where $\langle \alpha,\beta\rangle_{\xi,\eta}
=(\alpha,\xi)(\beta,\eta)-(\alpha,\eta)(\beta,\xi)$.
One sees that to  have a cancellation of the terms in the sum
corresponding to fixed rotation $s_\alpha s_\beta$, it is necessary
that, together with $\alpha$, $\beta$, the vector
$\gamma=s_\alpha(\beta)$, also belongs to $A$. This means
that $A$ is the root system $R$ for some Coxeter group.

To prove that $k$ is $G$-invariant, let us recall that if
it is so, then $[\nabla_\xi,\nabla_\eta]=0$ according
to Dunkl's theorem.
Suppose that $k$ is not invariant, i.e., there exist two
roots
$\alpha$, $\beta$ such that $k_\beta\neq k_\gamma$,
$\gamma=s_\alpha(\beta)$. Subtracting
from III the Dunkl identity with appropriate
coefficients, gives a relation of the form
\be
\sum_{\alpha,\beta\in A_+}
c_{\alpha,\beta}
\langle \alpha,\beta\rangle_{\xi,\eta}
\frac1{(\alpha,x)(s_\alpha(\beta),x)}\hat s_\alpha\hat s_\beta=0,
\ee
where $c_{\alpha,\beta}=0$, $c_{\gamma,\alpha}\neq 0$.
But this is impossible, since the pole at $(\gamma,x)=0$
cannot be canceled.
\end{proof}

Thus we have shown that Coxeter groups arise naturally
as a commutativity condition for a natural generalization
of Dunkl operators.

\subsubsection*{2. Functional equations for the
 coefficients of generalized
Dunkl operators}\label{s2}

Let us now consider for given Coxeter group $G$ the
following generalizations of Dunkl's operators:
\ben\label{e6}
\nabla_\xi=\partial_\xi+\sum_{\alpha\in R_+}
(\alpha,\xi)f_\alpha((\alpha,x))\hat s_\alpha.
\een
It is convenient to extend the definition of
$f_\alpha$ to all $\alpha\in R$, by setting
\ben\label{fa}
f_{-\alpha}(z)=-f_\alpha(-z).
\een
With this definition, $\nabla_\xi$ is independent
of the choice of the positive part $R_+$ of $R$.
In fact, we can replace $R_+$ by any subset of
$R$ consisting of one normal vector for each hyperplane,
without changing $\nabla_\xi$.

We may choose as before $R$ to consist of unit vectors,
but, since Weyl groups will be considered later,
 it is convenient to allow normal vectors to have arbitrary
length, still preserving the condition that we have two
normal vectors $\alpha$ and $-\alpha$ for each hyperplane
in $S$. At this point,
this is no generalization, since the operators
do not change if we replace $\alpha$ by a multiple
$\alpha'=c\alpha$ and replace the corresponding function
$f_\alpha$ by the function $f_{\alpha'}(z)=c^{-1}f_\alpha(c^{-1}z)$.

A calculation analogous to the previous one
leads to the following formula
\ben\label{e7}
[\nabla_\xi,\nabla_\eta]=
\sum_{\alpha,\beta\in R_+}
\langle\alpha,\beta\rangle_{\xi,\eta}
f_\alpha((\alpha,x))
f_\beta((s_\alpha(\beta),x))
\hat s_\alpha\hat s_\beta,
\een
where, as above, $\langle\alpha,\beta\rangle_{\xi,\eta}
=(\alpha,\xi)(\beta,\eta)-(\alpha,\eta)(\beta,\xi)$.

\begin{thm}
The commutativity condition for the operators \Ref{e6}
\be
[\nabla_\xi,\nabla_\eta]
\ee
is equivalent to the system of functional equations
\ben\label{e8}
\sum_{\alpha,\beta\in R_+:s_\alpha s_\beta=r}
\langle\alpha,\beta\rangle
f_\alpha((\alpha,x))
f_\beta((s_\alpha(\beta),x))
=0,
\een
for any given rotation $r$.
\end{thm}

Here $\langle\alpha,\beta\rangle$ denotes the oriented
area of the parallelogram with sides $\alpha$, $\beta$,
with respect to some orientation of the plane spanned
by these two vectors, which is perpendicular
to the rotation axis of $r$. Obviously, the equation \Ref{e8}
is independent of the choice of orientation.

\begin{example}
For the root system of type $A_2$ (see Fig.~\ref{a2}),
we have the following
functional equation for the three functions associated with
the three  roots labeled in Fig.~\ref{a2}. The functions
associated to the other roots are then determined by \Ref{fa}.
\ben\label{e9}
f(x-y)g(x-z)+g(y-z)h(y-x)+h(z-x)f(z-y)=0.
\een
\end{example}

\setlength{\unitlength}{40pt}
\begin{figure}[t]
\begin{picture}(8,3)(-4,-1.3)
\Root101\put(1.2,-0.05){\it f}
\Root35{0.5}
\Root{-3}5{0.5}\put(-0.7,1){\it g}
\Root{-1}01
\Root{-3}{-5}{0.5}\put(-0.7,-1){\it h}
\Root{3}{-5}{0.5}
\end{picture}
\caption{The root system of type $A_2$}\label{a2}
\end{figure}
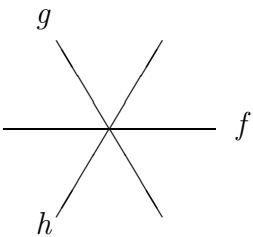

Let us now  discuss $G$-invariant solutions of
the functional equations \Ref{e9}: $f_{g(\alpha)}=
f_\alpha$ for all $g\in G$, or, equivalently,
\be
\hat g\nabla_\xi=\nabla_{g(\xi)}\hat g, {\rm\ for\ all\ } g\in G.
\ee
We  assume
that none of the functions $f_\alpha$ vanishes
identically.
We restrict ourselves
to the case where $G$ is the Weyl group $W$ a semisimple
Lie algebra. Obviously it is sufficient to consider
the case of a simple Lie algebra. Excluding the
one dimensional case $A_1$ where commutativity does
not give any restriction on $f_\alpha$, the first
case we consider is the case $A_{n-1}$, where $G=W$ is
the symmetric group $S_n$.

\begin{proposition}
For root systems of type $A_{n-1}$, $n\geq 3$, the only
$S_n$-invariant solution of \Ref{e8} meromorphic
in a neighborhood of the origin  is
\be
f_\alpha(z)=\frac Cz,
\ee
which corresponds to the usual Dunkl operators.
\end{proposition}

\begin{proof}
Consider the functional equation \Ref{e8}. For given
$r$, it involves only roots in the 2-plane orthogonal
to the rotation axis of $r$. This plane is spanned by
any two roots $\alpha$, $\beta$ with $s_\alpha s_\beta=r$.

In the $A_{n-1}$ case, the roots in any such plane build
a root system of type $A_1\times A_1$  or $A_2$.
In the former case roots are orthogonal and
\Ref{e8} is identically satisfied for all $f$.
It is therefore sufficient to consider the case $n-1=2$.
Let
us rewrite \Ref{e9} in the form
\ben\label{e10}
f(u)g(u+v)+g(v)h(-u)+h(-u-v)f(-v)=0.
\een
$S_3$-invariant solutions correspond to $f=g=h$, where
$f$ is an odd function, since $f_\alpha(z)=f_{-\alpha}(z)
=-f_\alpha(-z)$.
This leads to the following equation for
$f$:
\be
f(u)f(u+v)-f(v)f(u)+f(u+v)f(v)=0.
\ee
If $f$ vanishes at a point $u$, then $f(v)f(u+v)=0$ for
all $v$, and $f$ vanishes identically.
 If $f$ does not vanish anywhere,
we get,
after substitution $\phi=1/f$,
\be
\phi(u+v)=\phi(u)+\phi(v),
\ee
which has the only solution $\phi(z)=cz$.
\end{proof}

Let us consider some more examples of low rank. It will
be shown below that these examples essentially cover the
whole theory.

\begin{figure}[h]
\begin{picture}(8,4.5)(-4,-2.3)
\Root101 \put(1.2,-0.05){\it f}
\Root35{0.5}\put(1.75, 1){\it g}
\Root{-3}5{0.5}\put(0.7,1){\it f}
\Root{-1}01
\Root{-3}{-5}{0.5}\put(-0.6,1){\it f}
\Root{3}{-5}{0.5}\put(-1.7, 1){\it g}
\Root01{1.7}\put(-0.05,1.9){\it g}
\Root53{1.5}
\Root5{-3}{1.5}
\Root0{-1}{1.7}
\Root{-5}{-3}{1.5}
\Root{-5}{3}{1.5}
\end{picture}
\caption{The root system of type $G_2$}\label{g2}
\end{figure}
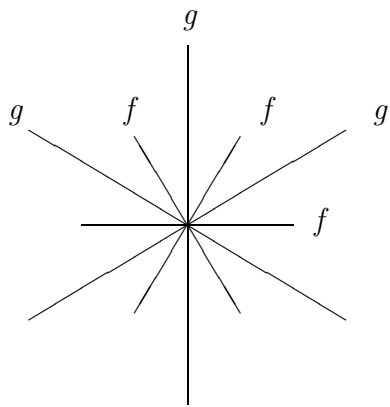

\begin{example}
In the case of $G_2$ (see Fig.~\ref{g2}), one has two
$A_2$ systems. Hence the the invariant solutions have
the form
\be
f(z)=\frac Az,\qquad g(z)=\frac Bz,
\ee
for arbitrary $A$ and $B$.
\end{example}

\begin{figure}[t]
\begin{picture}(8,2.5)(-4,-1.3)
\Root101\put(1.2,-0.05){\it f}
\Root111\put(1.2,1.2){\it g}
\Root011\put(0,1.2){\it f}
\Root{-1}11\put(-1.2,1.2){\it g}
\Root{-1}01
\Root{-1}{-1}1
\Root0{-1}1
\Root1{-1}1
\end{picture}
\caption{The root system of type $B_2$}\label{b2}
\end{figure}
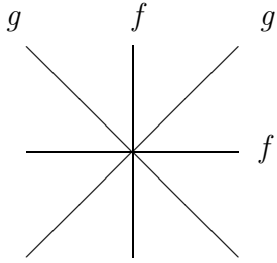

\begin{example}
In the $B_2$ case (see Fig.~\ref{b2}) one has the following
functional equation for the symmetric solution:
\ben\label{e11}
f(x)(g(x+y)+g(x-y))
+f(y)(g(x+y)-g(x-y))=0,
\een
where $f$ and $g$ are odd functions. In this
case we have more complicated solutions, such
as
\be
f(z)=\cot(z)
\qquad g(z)=
\frac1{\sin(z)}.
\ee
The general solution will be given below.
\end{example}

\begin{example}
In the $B_3$ case, $R=\{\pm e_i,\pm e_i\pm e_j\}$,
where $\{e_1,\, e_2,\, e_3\}$ is an orthonormal basis of $V$.
Let us find the most general invariant solution. The
roots $\alpha=e_1+e_2$, and $\beta=e_2+e_3$
form a subsystem isomorphic to $A_2$. Thus
$f_\alpha(z)=f_\beta(z)=C/z$. Now consider the
$B_2$ system generated by  $\gamma=e_1$ and $\alpha$.
So we have to solve the functional equation \Ref{e11}
with $g=C/z$:
\be
f_\gamma(x)(\frac C{x+y}+\frac C{x-y})
+
f_\gamma(y)(\frac C{x+y}-\frac C{x-y})=0,
\ee
or, equivalently (if $C\neq 0$),
\be
xf_\gamma(x)-yf_\gamma(y)=0,
\ee
which implies $f_\gamma(z)=D/z$. It follows
that the most general  invariant
solution  is Dunkl's solution.
\end{example}

\begin{thm}\label{T1}
For all root systems of simple Lie algebras
except $A_1$, $B_2$, all $W$-invariant solutions
of the functional equation \Ref{e8} have the form
\be f_\alpha(z)={k_\alpha\over z},
\ee
where $\alpha\mapsto k_\alpha$ is a $W$-invariant function on $R$.
\end{thm}

\begin{proof}
For all root systems except $A_1$, $B_2$, there exists
a subsystem isomorphic to $A_2$ (see  \cite{Bou})
All other two-dimensional subsystems are
isomorphic either to $A_2$ or $B_2$ (the
exceptional case of $G_2$ was considered above).
Continuing as in the $B_3$ case, we complete the proof.
\end{proof}

In the rest of this section we consider the invariant
$B_2$ case. In other words, we want to find the general
solution of the equation \Ref{e11}
\ben\label{ff}
f(x)(g(x+y)+g(x-y))
+f(y)(g(x+y)-g(x-y))=0,
\een
which are meromorphic in the vicinity of the
origin.

There are the following obvious solutions:
$f(z)\equiv 0$, $g$ an arbitrary function,
and $g(z)\equiv 0$, $f$ an arbitrary function.
We call these solutions {\em trivial}.

\begin{thm}\label{Tb2}
The general non-trivial solution of the functional
equation \Ref{ff} has the form
\be
g(z)=\frac A{\sn(\alpha z,k)}\quad,\qquad
f(z)= B(\log g(z))',
\ee
or, more explicitly
\ben\label{vv4}
g(z)=\frac A{\sn(\alpha z,k)}\quad,\qquad
f(z)=\frac B{\sn(\epsilon\alpha z,\tilde k)}\quad,
\een
where $\tilde k=(1-k)/(1+k)$, $\epsilon=-i(1+k)$, and
$A$, $B$ and $\alpha$ are arbitrary constants.
\end{thm}

\noindent Here $\sn(z, k)$ is the classical
Jacobi elliptic function (see, e.g.,  \cite{WW}).
In the degenerate cases we have the following
solutions:
\bea
f(z)&=&\frac A{\sin\alpha z}\quad,\qquad g(z)= B\cot\frac{\alpha z}2\quad,\\
f(z)&=&A\cot{\alpha z}\quad,\qquad g(z)=\frac B{\sin\alpha z}\quad,\\
f(z)&=&\frac Az\quad,\qquad g(z)= \frac Bz\quad,
\eea
with arbitrary constants $A$, $B$, $\alpha$. In particular,
we see that all non-trivial solutions are {\em odd} functions,
and thus lead to invariant Dunkl operators for root
systems of type $B_2$.

We proceed to prove the Theorem.

\begin{lemma}
If $f(z)$, $g(z)$ satisfy the equation
\Ref{ff}, then the same is true
for $\tilde f(z)$, $\tilde g(z)$, where
\ben\label{v5}
{\rm (i)}\qquad \tilde f(z)=\lambda f(\alpha z),
\qquad \tilde g(z)=\mu g(\alpha z),
\een
for arbitrary
constants $\lambda$, $\mu$, $\alpha$, or
\ben\label{v6}
{\rm (ii)} \qquad \tilde f(z)=g(z), \qquad
\tilde g(z)=f(z/2).
\een
\end{lemma}

\begin{proof}
The first symmetry is evident. To prove
the second one, it is sufficient to
change variables
\be
u=x+y,\qquad v=x-y.
\ee
Then equation \Ref{ff} takes the form
\be
g(u)
\left(
f\left(\frac{u+v}2\right)
+f\left(\frac{u-v}2\right)
\right)
+g(v)
\left(
f\left(\frac{u+v}2\right)
-f\left(\frac{u-v}2\right)
\right)=0.
\ee
\end{proof}

We will use these symmetries again and again,
in particular in the proof of the following

\begin{lemma}
There are no non-trivial solutions of \Ref{ff},
for which $f(z)$ or $g(z)$ is regular at the origin.
\end{lemma}

\begin{proof}
Because
of the symmetry \Ref{v6},
it is enough to consider only the case when $f(z)$ is regular
at the origin.
Putting $y=0$ in \Ref{ff}, we get $2f(x)g(x)=0$,
which means that the solution is trivial.
\end{proof}

\begin{lemma}
Non-trivial solutions of $\Ref{ff}$ are
odd functions:
\be
f(-z)=-f(z),\qquad g(-z)=-g(z).
\ee
\end{lemma}

\begin{proof}
Rewrite the equation \Ref{ff}
in the form
\be
\phi(y)\left(
g(x+y)+g(x-y)\right)
+\phi(x)\left(
g(x+y)-g(x-y)\right)=0,
\ee
where $\phi(z)=1/f(z)$ is regular at vanishes at $z=0$.
By putting $x=0$ in this relation, we obtain
\be
\phi(y)(g(y)+g(-y))=0
\ee
which implies for a non-trivial solution
\be
g(-y)=-g(y).
\ee
The fact that $f$ is odd follows then from the symmetry
\Ref{v6}.
\end{proof}

Let us introduce $\lambda=\phi'(0)$, where
$\phi(z)=1/f(z)$, as before.

\begin{lemma} If $f(z)$, $g(z)$ is a non-trivial
solution of \Ref{ff}, then
\bean
 & {\rm (i)}& \lambda =\phi'(0)\neq 0, \nonumber\\
\label{v7} & {\rm (ii)}& \frac{ g'(x)}{g(x)}=-\lambda f(x),\\
\label{v8}  &{\rm (iii)}&
\lambda
f(x+y)=
\frac
{f'(y)f(x)-f'(x)f(y)}
{f^2(x)-f^2(y)}.
\eean
\end{lemma}

\begin{proof}
Rewrite equation \Ref{ff} in the form
\ben\label{v9}
g(x+y)=\frac{f(y)-f(x)}{f(y)+f(x)}g(x-y).
\een
Taking the logarithm of both sides and applying
the operator $\partial_x+\partial_y$, gives
\bean\label{v10}
\frac{g'(x+y)}{g(x+y)}
&=&\frac{f(x)f'(y)-f(y)f'(x)}
{f^2(y)-f^2(x)}\\
\label{v10a} &=&\frac{\phi'(x)\phi(y)-\phi'(y)\phi(x)}
{\phi^2(x)-\phi^2(y)}\quad.
\eean
Putting $y=0$ in \Ref{v10a} implies
\be
\frac{g'(x)}{g(x)}=-\lambda
f(x).
\ee
In particular, if $\lambda=0$, $g$ is constant and therefore
regular, which is impossible. Now the formula \Ref{v8}
follows from \Ref{v7} and \Ref{v10}
\end{proof}

Using the symmetry \Ref{v5}, \Ref{v6}, we may set without
loss of generality $\lambda=1$, so that
\be
f(z)=\frac 1z +O(z),\qquad g(z)=\frac 1z+O(z),\qquad (z\to 0).
\ee
The function $f$ satisfies the functional equation
(addition theorem)
\ben\label{v13}
f(x+y)=
\frac{f(x)f'(y)-f(y)f'(x)}
{f^2(x)-f^2(y)}.
\een
Rewrite it in the following form
\ben\label{v14}
f(x+y)=
\frac
{\phi'(y)f(x)+f'(x)\phi(y)}
{1-\phi^2(y)f^2(x)},
\een
and expand the right hand side near $y=0$, using
$\phi(z)=z+az^3+O(z^5)$:
\be
f(x+y)=f(x)+f'(x)y+(3a\,f(x)+f^3(x))y^2+O(y^3).
\ee
By comparing this with the Taylor expansion,
one has
\ben\label{v15}
 f''(x)=2f^3(x)+6a\,f(x),
\een
implying, after multiplication by $f'$ and integration, that
\ben\label{v16}
(f')^2=f^4+6af^2+b,
\een
for some constant $b$. The function $\phi=1/f$ is thus
a regular odd solution of the equation
\ben\label{v17}
(\phi')^2=1+6a\phi^2+b\phi^4,
\een
and therefore coincides with the Jacobi
elliptic function
\ben\label{v18}
\phi(x)=\frac{\sn(\epsilon x, k)}\epsilon,
\een
where $(1+k^2)\epsilon^2=-6a$, $k^2\epsilon^4=b$.
Recall that $\sn$ is the solution of
the equation $(s')^2=(1-s^2)(1-k^2s^2)$, with
initial condition $s(0)=0$.
It satisfy the addition formula discovered by
A. Cayley (see \cite{WW})
\be
s(x+y)=
\frac{s^2(x)-s^2(y)}
{s(x)s'(y)-s(y)s'(x)}.
\ee
This implies the relation \Ref{v13} for
$f(x)=\epsilon/\sn(\epsilon x,k)$. Note that
by \Ref{v6}, also $g$ has the same form as $f$,
in general with different values of $\epsilon$ and
$k$.

\begin{lemma}
If $f(x)=\epsilon/\sn(\epsilon x,k)$ and
$g'(x)/g(x)=-f(x)$ then $f(x)$, $g(x)$ are
a non-trivial solution of the functional equation
\Ref{ff}.
\end{lemma}

\begin{proof}
We have
\bea
2\,\frac{g'(x+y)}{g(x+y)}
&=&
-2\,f(x+y)\\
 &=& 2\,\frac{f'(x)f(y)-f'(y)f(x)}{f^2(x)-f^2(y)}\\
 &=&\frac{f'(y)-f'(x)}{f(y)-f(x)}-
\frac{f'(y)+f'(x)}{f(y)+f(x)}.
\eea
So
\be
(\partial_x+\partial_y)\log g(x+y)
=
(\partial_x+\partial_y)\log\frac{f(y)-f(x)}{f(y)+f(x)}.
\ee
This means that
\be
g(x+y)=\frac{f(y)-f(x)}{f(y)+f(x)}\psi(x-y),
\ee
for some function $\psi$. Rewriting this equation
as
\be
g(x+y)=
\frac{\phi(x)-\phi(y)}{\phi(x)+\phi(y)}\psi(x-y),
\qquad \phi=1/f,
\ee
and putting $y=0$, gives $\psi(x)=g(x)$, and therefore
\be
g(x+y)=\frac{f(y)-f(x)}{f(y)+f(x)}g(x-y),
\ee
which is equivalent to \Ref{ff}.
\end{proof}

To finish the proof of Theorem \ref{Tb2},
we have to prove that the relation between
$k$, $\tilde k$, $\epsilon$ in
\be
g(x)=\frac1{\sn(x,k)},\qquad
f(x)=\frac\epsilon{\sn(\epsilon x,\tilde k)}\quad,
\ee
where $g'/g=-f$ is the one given in
Theorem \ref{Tb2}.

The function $1/f=-g/g'=\sn(\epsilon x,\tilde k)/\epsilon$
satisfies the equation
\ben\label{vv20}
[(g/g')']^2
=
[1-\epsilon^2(g/g')^2]
[1-\tilde k^2\epsilon^2(g/g')^2],
\een
or, equivalently,
\be
\left(
(g')^2-gg''
\right)^2
=\left((g')^2-\epsilon^2g^2\right)
\left((g')^2-\tilde k^2\epsilon^2g^2\right).
\ee
Substituting $(g')^2=(g^2-1)(g^2-k^2)$,
$g''=2g^3-(k^2+1)g$, into \Ref{vv20}, gives
the following two possibilities
\begin{enumerate}
\item[(i)]
$2k=-(k^2+1)-\epsilon^2$, $-2k=-(k^2+1)-\tilde k^2\epsilon^2$
\item[(ii)]
$-2k=-(k^2+1)-\epsilon^2$, $2k=-(k^2+1)-\tilde k^2\epsilon^2$
\end{enumerate}
In the first case, we get $\tilde k=(1-k)/(1+k)$,
$\epsilon=i(k+1)$, as required. The second case leads
to an equivalent answer. Theorem \ref{Tb2} is proved.

\par\vspace{.5\baselineskip}

\noindent{\em Remark.} The transformation of elliptic
functions  $g\to f$ is of
second order and therefore can be reduced to the well-known Landen
transformation (see \cite{WW}).
One can check that it is the composition of the
``imaginary Jacobi transformation'', which is the unimodular
transformation with the action on the homology of the curve
described by the matrix $J$
  \be
J=\left(\begin{array}{cc}0 & 1\\-1& 0\end{array}\right).
\ee
and  Landen's transformation with  matrix $L$,
   \be
L=\left(\begin{array}{cc}2 & 0\\0& 1\end{array}\right).
\ee
  Let us call it LJ-transformation. Thus the functional equation
\Ref{ff}
describes a pair of elliptic functions related by the
LJ-transformation. This fact has found recently an interesting
topological application in the theory of elliptic genera
\cite{BuVe}.

Note that $f$ and $g$, as elliptic functions of second order
 ``live'' on different elliptic curves,
one of which is a double cover of the other. In particular,
the solutions found here are not special cases of the
elliptic solutions considered in the next Section.

\subsubsection*{3. Elliptic Dunkl operators}
In this section we consider only the
case where $R$ is the root system
of a semisimple Lie algebra with
Weyl group $G=W$.
 Let us consider the elliptic curve
with modular parameter $\tau$,  Im$(\tau)>0$,
and the family of functions
\ben\label{sigma}
\sigma_\lambda(z)=\frac
{\theta_1(z-\lambda)\theta_1'(0)}
{\theta_1(z)\theta_1(-\lambda)}, \qquad \lambda\in\C\setminus
\Z+\tau\Z,
\een
given in terms of Jacobi's theta function
\be
\theta_1(z)=-\sum_{n=-\infty}^{\infty}
e^{2\pi i(z+\frac12)(n+\frac12)+\pi i\tau(n+\frac12)^2}.
\ee
The functions $\sigma_\lambda$ have the following
defining properties:
\begin{enumerate}
\item[(i)] $\sigma_\lambda(z+1)=\sigma_\lambda(z)$.
\item[(ii)] $\sigma_\lambda(z+\tau)=
e^{2\pi i\lambda}\sigma_\lambda(z)$.
\item[(iii)] $\sigma_\lambda$ is meromorphic, its poles
are on the lattice $\Z+\tau \Z$, and $\sigma_\lambda(z)
=1/z+{O}(1)$ as $z\to 0$.
\end{enumerate}
More properties of this functions are given in the Appendix.
\begin{thm}\label{tfa}
For any generic $\lambda\in V_\C=V\otimes_\R\C$,
and $W$-invariant function
$k_\alpha$,
the functions
\be f_\alpha(z)=k_\alpha\sigma_{(\alpha^\vee,\lambda)}(z),\ee
where $\alpha^\vee=2\alpha/(\alpha,\alpha)$, satisfy
the functional equations \Ref{e8}.
\end{thm}

\begin{proof}
Fix the rotation $r$. All roots involved in the left hand
side of the functional equation \Ref{e8}
\ben\label{I}
I(x)=\sum_{\alpha,\beta\in R_+:s_\alpha s_\beta=r}
\langle\alpha,\beta\rangle
\sigma_{(\alpha^\vee,\lambda)}((\alpha,x))
\sigma_{(\beta^\vee,\lambda)}((s_\alpha(\beta),x))
\een
lie on the same two-dimensional plane. Consider $I(x)$
as a meromorphic function of $x\in V_\C$,
and let $P^\vee=\{p\in V\,|(p,\alpha)\in\Z\; \forall\alpha\in R\}$.
Then if $p\in P^\vee$, $I(x+p)=I(x)$, and as
$x\to x+p\tau$, the term labeled by $(\alpha,\beta)$ in
the sum \Ref{I} gets multiplied by
\be
e^{2\pi i\left( (\alpha^\vee,\lambda)(\alpha,p)
+(\beta^\vee,\lambda)(s_\alpha(\beta),p)\right)}.
\ee
Since $s_\alpha(\lambda)=\lambda-(\alpha^\vee,\lambda)\alpha$,
we see that the multiplier can be rewritten as
\be
e^{2\pi i(\lambda-s_\alpha s_\beta(\lambda),p)},
\ee
and is therefore the same for all terms  in the sum. It
follows that $I(x)$ has the quasi-periodicity property
\ben\label{per}
I(x+q+p\tau)=
e^{2\pi i(\lambda-r(\lambda),p)}
I(x), \qquad q+p\tau\in P^\vee+\tau P^\vee.
\een
Let us now consider the poles of the function $I$.
Poles may
appear when $x$ is on the hyperplanes $(\alpha,x)=0$,
$\alpha\in R_+$, or
their translates by $P^\vee+\tau P^\vee$.
If $x$ approaches the hyperplane $(\alpha,x)=0$,
the singular terms in the sum \Ref{I} are the
terms indexed by $\alpha,\beta$ and $\gamma,\delta$ where
$s_\gamma(\delta)=\pm\alpha$. As all roots are on a plane
this implies that $\gamma=\mp \beta$. The sign is $+$
(thus $s_\gamma(\delta)=-\alpha$) since $\gamma>0$.
In particular only two terms are singular in \Ref{I}.
The coefficient of the (simple) pole is (cf.\ (iii) above)
\be
\langle\alpha,\beta\rangle
\sigma_{(\beta^\vee,\lambda)}
((s_\alpha(\beta),x))
-
\langle\beta,\delta\rangle
\sigma_{(\beta^\vee,\lambda)}
((\beta,x)).
\ee
This expression vanishes on the hyperplane $(\alpha,x)=0$
because $s_\alpha(x)=x$ there, and
\be
\langle \beta,\delta\rangle=
-\langle s_\beta(\beta),s_\beta(\delta)\rangle
=-\langle\beta,\alpha\rangle=
\langle\alpha,\beta\rangle.
\ee
It follows that the singularity at $(\alpha,x)=0$ (and thus on
all affine hyperplanes $(\alpha,x)=n+m\tau$, $n$, $m\in \Z$
by \Ref{per}) is removable. We conclude that $I$ has no
singularity on $V_\C$, and has the quasi-periodicity property
\Ref{per}. It therefore vanishes, by Fourier series theory.
\end{proof}

\begin{corollary}
The operators
\be
\nabla^\lambda_\xi=\partial_\xi+\sum_{\alpha\in R_+}
k_\alpha(\alpha,\xi)
\sigma_{(\alpha^\vee,\lambda)}((\alpha,x))\hat s_\alpha,
\ee
form a commutative family:
\be
[\nabla^\lambda_\xi,\nabla^\lambda_\eta]=0.
\ee
\end{corollary}
Let us call these operators {\em elliptic Dunkl operators}.
They are not $W$-invariant but $W$-equivariant
\ben
\label{equi}
\hat w\nabla_\xi^\lambda\hat w^{-1}
=
\nabla^{w(\lambda)}_{w(\xi)}.
\een

\begin{example}
In the $A_{n-1}$ case, we identify functions on
$V=\{\R^n\,|\, \Sigma_ix_i=0\}$ with functions
$f$ on $\R^n$ such that $f(x_1+a,\dots,x_n+a)$
is independent of $a\in\R$. Let $e_1,\dots,e_n$
be the standard basis of $\R^n$ and put
\be
\bar e_i=e_i-\frac1n\sum_{j=1}^ne_j\in V.
\ee
Then Dunkl operators are linear combinations of
the commuting operators $\nabla^\lambda_i=\nabla^\lambda_{\bar e_i}$:
\ben\label{edo}
\nabla_i^\lambda=
\frac\partial{\partial x_i}+k\sum_{j:j\neq i}
\sigma_{\lambda_i-\lambda_j}(x_i-x_j)\hat s_{ij},
\een
where $\hat s_{ij}$ is the operator that interchanges the
$i$th and $j$th variable.
\end{example}

\par\vspace{.5\baselineskip}

\noindent{\em Remark.} It is interesting to note that
the usual Dunkl operator reminds Moser's $L$-matrix
for the Calogero problem \cite{Mo}, whereas the elliptic
Dunkl operator \Ref{edo} reminds Krichever's generalization
\cite{Kr}, see also \cite{OP1}.
The general solution of the corresponding
functional equation (different from ours) was found
by Bruschi and Calogero \cite{BC}. More general functional
equations motivated by these problems, as well as
some topological problems, were introduced and solved by
one of the authors \cite{Bu}, \cite{Bu2}.

\subsubsection*{4. General solution of the functional equation in the
$A_{n-1}$ case}

In this section we show that elliptic Dunkl operators
are essentially the only solutions of the functional
equation \Ref{e8}, in the $A_{n-1}$ case, $n-1\geq 2$.

As before, we start with $A_2$. In this case, the
functional equation is \Ref{e10}
\ben\label{f}
f(u)g(u+v)+g(v)h(-u)+h(-u-v)f(-v)=0.
\een
We must find the most general solutions of this
functional equation, where $f$, $g$, and $h$ are assumed
to be meromorphic functions defined in a neighborhood
of the origin.

First of all, if one of the three functions vanishes identically,
then the functional equation says that the product of the
other two vanishes, and we get a solution where two functions
vanish and the third is arbitrary. We call these solution
trivial, and consider from now on only solutions where
$f$, $g$ and $h$ are not identically zero.

\begin{lemma}\label{le1}
If $f$, $g$, $h$ satisfy \Ref{f}, then
{\rm (i)}
$\tilde f(z)=g(z)$,
$\tilde g(z)=h(z)$,
$\tilde h(z)=f(z)$,
{\rm (ii)}
$\tilde f(z)=af(bz)e^{\alpha z}$,
$\tilde g(z)=ag(bz)e^{\beta z}$,
$\tilde h(z)=af(bz)e^{\gamma z}$,
for arbitrary constants $a,\dots,\gamma$ such that
$\alpha+\beta+\gamma=0$,

\noindent also satisfy \Ref{f}.
\end{lemma}

\begin{proof}
 Replacing $(u,v)$ by $(-u-v,u)$ in \Ref{f} implies (i).
Property (ii) is easy to check.
\end{proof}

\begin{proposition}\label{pr2}
The only non-trivial solutions of \Ref{f} holomorphic around
the origin are
\be
f(u)=ae^{\alpha u},\qquad
g(u)=be^{\beta u},\qquad
h(u)=ce^{\gamma u},
\ee
where $a$, $b$, $c\neq 0$,
$ab+bc+ac=0$, $\alpha+\beta+\gamma=0$.
\end{proposition}

\begin{proof} It is easy to check that these are solutions.
We prove uniqueness. Suppose $f$, $g$, $h$ are a non-trivial solution
of \Ref{f}, defined and holomorphic in a neighborhood of the
origin. Taking $v=0$ in \Ref{f}, we get
\ben\label{p1}
f(u)g(u)+(f(0)+g(0))h(-u)=0.
\een
We have $f(0)+g(0)\neq 0$ since $fg$ does not vanish identically.
Similarly, by Lemma \ref{le1},
\bean
\label{p2}g(u)h(u)+(g(0)+h(0))f(-u)&=&0,\\
\label{p3}h(u)f(u)+(h(0)+f(0))g(-u)&=&0.
\eean
Introduce the functions $F(u)=f(u)f(-u)$,
$G(u)=g(u)g(-u)$, $H(u)=h(u)h(-u)$, and
$S(u)=f(u)g(u)h(u)$. Multiplying \Ref{p1},
\Ref{p2}, \Ref{p3} by $h(u)$, $f(u)$ and
$g(u)$, respectively, we obtain
\ben\label{p4}
S(u)=\lambda F(u)=\mu G(u)=\nu H(u),
\een
for some non-zero constants $\lambda$, $\mu$, $\nu$.
In particular $S$ is an even function, and thus
\be
S(u)^2=S(u)S(-u)=F(u)G(u)H(u).
\ee
Hence $F(u)^3={\rm const}F(u)$, and $F$ is a constant. Similarly,
$G$ and $H$ are constant functions.
Thus we have
\be
f(u)f(-u)=a^2,\qquad
g(u)g(-u)=b^2,\qquad
h(u)h(-u)=c^2,
\ee
for some constants $a$, $b$, $c\neq 0$.
We can therefore write
\ben\label{p6}
f(u)=ae^{\phi(u)},\qquad g(u)=be^{\psi(u)},
\qquad h(u)=ce^{\eta(u)},
\een
for some odd functions $\phi$, $\psi$, $\eta$, holomorphic
in a neighborhood of $0$.
Since $S(u)=f(u)g(u)h(u)$ is constant, we have
$\phi+\psi+\eta=0$.
Inserting \Ref{p6}
in the functional equation gives
\be
ab\,e^{\phi(u)+\psi(u+v)}
+bc\,e^{\psi(v)-\eta(u)}
+ac\,e^{-\eta(u+v)-\phi(v)}
=0,
\ee
which, after elimination of $\eta=-\phi-\psi$ can be
recast in the more convenient form
\ben\label{p7}
ab+bc\,\Psi(u,v)+ac\,\Phi(u,v)^{-1}=0,
\een
with $\Phi(u,v)=\exp(\phi(u+v)-\phi(u)-\phi(v))$, and
$\Psi(u,v)=\exp(\psi(u+v)-\psi(u)-\psi(v))$. In particular,
setting $u=v=0$, we obtain the condition $ab+bc+ac=0$.
Moreover, we have $\Phi(-u,-v)=\Phi(u,v)^{-1}$, and
$\Psi(-u,-v)=\Psi(u,v)^{-1}$, since $\phi$ and $\psi$
are odd functions. Replacing $(u,v)$ by $(-u,-v)$ in
\Ref{p7} yields the equation
\ben\label{p8}
ab+bc\,\Psi(u,v)^{-1}+ac\,\Phi(u,v)=0.
\een
Elimination of $\Psi$ from \Ref{p7}, \Ref{p8} gives
a non trivial quadratic equation with constant coefficients
for $\Phi$. Thus $\Phi$ is constant, implying that
$\Psi$ is constant as well. we conclude that
$\phi(u+v)=\phi(u)+\phi(v)$, and $\psi(u+v)=\psi(u)+\psi(v)$,
which leaves us with the solution $\phi(u)=\alpha u$,
$\psi(u)=\beta u$.
\end{proof}

{}From now on, we consider the case when $f(u)$ has a
pole of order $p>0$ at the origin. It is easy to see that
$h$ and $g$ also have a pole of the same order $p$ at
$u=0$.

Inserting the expansion
\bean
f(u)&=&\frac a{u^p}+\cdots,\qquad\\
g(u)&=&\frac b{u^p}+\cdots,\qquad\\
h(u)&=&\frac c{u^p}+\cdots,
\eean
into \Ref{f}, and multiplying the relation by
$u^p(u+v)^pv^p$, yields a series in $u$ and $v$
which starts as
\be
ac\, u^p+ ab\,v^p+(-1)^pbc\,(u+v)^p.
\ee
We see that $p=1$ is the only case in which cancellation is possible.
In this case,
\be
ac-bc=0,\qquad ab-bc=0,
\ee
which implies $a=b=c$. Without loss of generality,
we consider the case $a=b=c=1$.

Suppose that $f$, $g$, $h$ are a solution of \Ref{f} with
a simple pole at the origin with unit residue:
\bean\label{o1}
f(u)&=&\frac1u+f_0+\cdots,\\
g(u)&=&\frac1u+g_0+\cdots,\\
h(u)&=&\frac1u+h_0+\cdots .
\eean
The left hand side of \Ref{f} has a Laurent expansion
at $v=0$:
\be
f(u)g(u)+(\frac1v+g_0)h(-u)
+(h(-u)-vh'(-u))(-\frac1v+f_0)+O(v).
\ee
The constant term gives the equation
\ben
\label{o2}f(u)g(u)+(f_0+g_0)h(-u)+h'(-u)=0.
\een
Similarly, by Lemma \ref{le1} (i),
\bean
\label{o3}g(u)h(u)+(g_0+h_0)f(-u)+f'(-u)&=&0,\\
\label{o4}h(u)f(u)+(h_0+f_0)g(-u)+g'(-u)&=&0.
\eean
Introduce $F(u)=f(u)f(-u)$,
$G(u)=g(u)g(-u)$, $H(u)=h(u)h(-u)$, and
$S(u)=f(u)g(u)h(u)$, as above. Note that
$F$, $G$ and $H$ are even functions.
Then $F'(u)=f'(u)f(-u)-f'(-u)f(u)$, and \Ref{o3}
implies that $F'(u)=S(u)-S(-u)$. More generally,
(\ref{o2}--\ref{o4}) imply
\be
F'(u)=G'(u)=H'(u)=S(u)-S(-u).
\ee
We can then write
\ben\label{abc}
F(u)=a-P(u), \qquad
 G(u)=b-P(u),\qquad H(u)=c-P(u),
\een
where $P(u)=1/u^2+O(u^2)$ has no constant term,
and $a$, $b$, $c$ are integration constants. The
even function $P$ obeys
\ben\label{o8}
P'(u)=-S(u)+S(-u).
\een
Let us compute the derivative
$S'=(f'/f+g'/g+h'/h)S$ using (\ref{o2}--\ref{o4})
\bea
S'(u)&=&\alpha S(u)-G(u)H(u)-H(u)F(u)-F(u)G(u)
\\
 &=&\alpha S(u)-3P(u)^2+\beta P(u)+\gamma,
\eea
for some constants $\alpha$, $\beta$, $\gamma$.
Let us consider the even and odd part of this equation
separately:
\bean
\label{o9}
\frac d{du}(S(u)+S(-u))&=&\alpha(S(u)-S(-u))=-\alpha P'(u)\\
\label{o10}
\frac d{du}(S(u)-S(-u))&=&\alpha(S(u)+S(-u))
-6P(u)^2+2\beta P(u)+2\gamma.
\eean
Integrating \Ref{o9} gives
\ben\label{o9a}
S(u)+S(-u)=-\alpha P(u)+\delta,
\een
for some $\delta$.
By subtracting \Ref{o8} from this equation, we obtain
\ben\label{o11}
S(u)=\frac12(-\alpha P(u)+\delta-P'(u)).
\een
Inserting \Ref{o9a} and \Ref{o8}
into \Ref{o10}
yields finally the differential equation

\be
P''(u)=6P(u)^2+(\alpha^2-2\beta) P(u)-\alpha\delta-2\gamma.
\ee
Let us multiply this equation by $P'(u)$ and integrate.
We get
\be
P'(u)^2=4P(u)^3+c_1P(u)^2+c_2 P(u)+c_3.
\ee
In fact $c_1=0$, since, by construction,
the Laurent expansion of $P$ has no constant term.
This equation is well-known to have a unique meromorphic
solution with double pole at the origin. It is
the Weierstrass function $\wp$ for some elliptic
curve determined by the coefficients $c_2$, $c_3$,
or one of its degenerations $\omega^2(1/\sin(\omega u)^2-1/3)$,
$1/u^2$ (see \cite{WW}).
After rescaling of the variables as in Lemma \ref{le1}
if necessary, we may assume that the periods of $\wp$
are $1$ and $\tau$, and that $\omega=\pi$.
By writing the constants $a$, $b$, $c$ in
\Ref{abc}
as $P(\lambda)$, $P(\mu)$, $P(\nu)$ respectively (this is
always possible since $P$ defines a surjective map
from the elliptic curve
 $\C/\Z+\tau\Z$, or, in the degenerate cases, $(\C/\Z)\cup\{i\infty\}$,
$\C\cup\{i\infty\}$,
 onto the Riemann sphere), we see that the equation
$f(u)f(-u)=P(\lambda)-P(u)$ has the solution (see the Appendix)
\be
f(u)=\sigma_\lambda(u),
\ee
and, in the degenerate cases,
\bea
f(u)&=&\frac{\pi\sin(\pi(u-\lambda))}
{\sin(\pi u)\sin(-\pi\lambda)}=\pi(\cot(\pi u)-\cot(\pi\lambda)),\\
f(u)&=&\frac1u-\frac1\lambda.
\eea
The general solution in a neighborhood of the origin
is $f(u)\exp\phi(u)$ for some odd
function $\phi(u)$ regular at the origin.
 Similar formulas hold for $g$ and $h$: the functions
\be
f(u)=\sigma_\lambda(u),\qquad
g(u)=\sigma_\mu(u),\qquad
h(u)=\sigma_\nu(u),
\ee
and their degenerations, are a solution of the functional
equation, provided that $\lambda+\mu+\nu=0$ (modulo the lattice).
We now show that these are the only (up to transformations
of Lemma \ref{le1} (ii))
solutions such that \Ref{abc} holds
with $a=P(\lambda)$, $b=P(\mu)$, $c=P(\nu)$.
Suppose $\tilde f$, $\tilde g$, $\tilde h$ is another solution. Then
\ben\label{o19}
\tilde f(u)=e^{\phi(u)}f(u),\qquad
\tilde g(u)=e^{\psi(u)}g(u),\qquad
\tilde h(u)=e^{\eta(u)}h(u),
\een
for some odd functions $\phi$, $\psi$, $\eta$. Since the product
$S=fgh$ is expressed in terms of $P$ (see \Ref{o11}), the
sum $\phi+\psi+\eta$ vanishes identically. Inserting
\Ref{o19} in \Ref{o3}, \Ref{o4}, we deduce immediately
that $\phi'(u)=\psi'(u)=0$ for all $u$.
Thus $\phi$ and $\psi$ are linear functions.

Combining this result with Prop.\ \ref{pr2}, we
obtain the following theorem.

\begin{thm}\label{Ta2}
The following list exhausts all non-trivial solutions
of the functional equation
\be
f(u)g(u+v)+g(v)h(-u)+h(-u-v)f(-v)=0.
\ee
\noindent Elliptic solutions:
\be
f(u)=a\sigma_\lambda(bu)e^{\alpha u},\qquad
g(u)=a\sigma_\mu(bu)    e^{\beta u},\qquad
h(u)=a\sigma_\nu(bu)    e^{\gamma u},
\ee
where  the function
$\sigma_\lambda$ is defined in \Ref{sigma}.

\noindent Trigonometric solutions:
\bean\label{deg}
f(u)=\frac{a\, \sin(b(u-\lambda))}
{\sin(bu)\sin(-b\lambda)}e^{\alpha u},& &
g(u)=\frac{a\, \sin(b(u-\mu))}
{\sin(bu)\sin(-b\mu)}e^{\beta u},\\
h(u)&=&\frac{a\, \sin(b(u-\nu))}
{\sin(bu)\sin(-b\nu)}e^{\gamma u}.\nonumber
\eean
Rational solutions:
\ben\label{ra}
f(u)=(\frac au-\frac1\lambda)e^{\alpha u},\qquad
g(u)=(\frac au-\frac1\mu)e^{\beta u},\qquad
h(u)=(\frac au-\frac1\nu)e^{\gamma u}.
\een
Regular solutions:
\ben\label{re}
f(u)=-\frac1\lambda e^{\alpha u},\qquad
g(u)=-\frac1\mu e^{\beta u},\qquad
h(u)=-\frac1\nu e^{\gamma u}.
\een
The parameters in these solutions are complex
numbers $a$, $b$, $\alpha$, $\beta$, $\gamma$,
$\lambda$, $\mu$, $\nu$ satisfying
$\alpha+\beta+\gamma=0$, $\lambda+\mu+\nu=0$.

In the trigonometric and rational cases, the limiting
cases where $\lambda$, $\mu$, $\nu$ take the value
$\pm i\infty$ are permitted.
\end{thm}

\noindent{\em Remark.} Notice that all these solutions
are limiting cases (degenerations) of the elliptic solutions.

This theorem covers the case $A_2$, but gives
immediately the answer in the case $A_{n-1}$, $n-1\geq 2$.
In this case, we identify, as above, functions
on $V$ with translation invariant functions on
$\R^n$.

\begin{corollary} Let $n\geq 3$. The $n$ operators
\be
\nabla_i=\partial_i+\sum_{j:j\neq i}f_{ij}(x_i-x_j)\hat s_{ij},
\qquad i=1,\dots,n,
\ee
with $f_{ij}\not\equiv 0$, $f_{ij}(u)=-f_{ji}(-u)$
are pairwise commutative iff
$f_{ij}(x)=k\,\sigma_{\lambda_i-\lambda_j}(bx)
e^{(\alpha_i-\alpha_j)x}$
where $\sigma_\lambda$ is
the function defined in \Ref{sigma}, or one of its
degenerations (see above).
\end{corollary}

\subsubsection*{5. Quantum Dunkl operators}

Fix complex parameters $\tau$, $\mu$ and $\kappa$ such
that Im$(\tau)>0$, $\mu\not\in\Z+\tau\Z$ and $\kappa\neq 0$.
Consider the operator $R(\lambda)$, depending on the ``spectral
parameter'' $\lambda\in\C$, and acting on the space of,
say, meromorphic functions of two complex variables $x_1$
and $x_2$:
\be
R(\lambda)f(x_1,x_2)=
\frac1{\sigma_\mu(\lambda)}
 \{\sigma_\mu(x_{12}+{\scriptstyle\frac\mu\kappa})
f(x_1+{\scriptstyle\frac\mu\kappa},x_2-{\scriptstyle\frac\mu\kappa})
-
\sigma_\lambda(x_{12}+{\scriptstyle\frac\mu\kappa})
f(x_2,x_1)\}.
\ee
We use the notation $x_{12}$ to denote the difference
$x_1-x_2$.
The operator $R$ obeys the quantum Yang--Baxter equation
\be
\RR 12\RR 13\RR 23=\RR 23\RR 13\RR 12.
\ee
The two sides of this equation are operators
acting on functions of three variables, and the
notation $R(\lambda)^{(ij)}$ indicates the operator
$R(\lambda)$ acting on a function of several variables,
by viewing it as a function of the $i$th and $j$th
variable. This solution of the Yang--Baxter equation
is essentially the one introduced in \cite{FePa} as three-parameter
generalization of the two-parameter solution of
Shibukawa and Ueno \cite{ShiUe}. For positive integer
values of $\kappa$ it admits a restriction to a finite
dimensional
subspace  coinciding with Belavin's solution \cite{Be}.

With the normalization used here we have ``unitarity''
\be\RR 12\RR 21=\rm{Id},\ee and $R(0)f(x_1,x_2)=f(x_2,x_1)$.
Moreover $R(\lambda)$ tends to the identity as $\mu$
goes to zero.

Solutions of the quantum Yang--Baxter equation
with these properties can be used to construct
commuting operators $T_i(\lambda_1,\dots,\lambda_n)$
(related to transfer matrices of integrable models
of statistical mechanics):
\ben\label{ti}
T_i(\lambda_1,\dots,\lambda_n)=
R^{(i,i+1)}\cdots R^{(i,n)}R^{(i,0)}\cdots R^{(i,i-1)}.
\een
These operators act on functions of $n$ variables, and,
as before, $R^{(ij)}=R(\lambda_i-\lambda_j)^{(ij)}$ is
the operator $R(\lambda_i-\lambda_j)$
acting on a function of $n$ variables
by viewing it as a function of  the $i$th and $j$th variable.

Our result is that elliptic Dunkl operator $\nabla_i^\lambda$
defined in \Ref{edo} can be
obtained as semiclassical limit of the quantum operators,
if $k$ is integer.

\begin{thm} Let $T_i(\lambda_1,\dots,\lambda_n)$ be
the operators \Ref{ti} acting on functions $f$ on $\R^n$,
such that $f(x_1+a,\dots,x_n+a)=f(x_1,\dots,x_n)$ for all
$a\in \R$.
For all integer $k$, we have
\be
T_i(\lambda_1,\dots,\lambda_n)={\rm Id}+\frac nk\mu
g^{-1}_{\kappa/n}\nabla_i^\lambda
g_{\kappa/n}+O(\mu^2),
\ee
where $g_m$ is the function
\be
g_m(x)=
\prod_{i<j}\theta_1(x_{ij})^m
\exp(m\sum_{i\neq j}x_i
\theta'_1(\lambda_{ij})/\theta_1(\lambda_{ij})
),
\ee
viewed as multiplication operator, and the parameter
$\kappa$ of $R$ is given by $\kappa=(-1)^knk$.
\end{thm}

\begin{proof} Let $\rho(x)=\theta_1'(x)/\theta_1(x)$.
We first compute the expansion of $R$ to first order.
\be
R(\lambda)={\rm Id}+\mu\, r(\lambda)+O(\mu^2),
\ee
where $r$, the ``classical $r$-matrix'', is
the differential-difference operator
\be
r(\lambda)=
\frac 1\kappa(\partial_1-\partial_2)+
\sigma_\lambda(x_{12})\hat s_{12}
+\rho(\lambda)-\rho(x_{12}).
\ee
This implies that
\be
T_i(\lambda_1,\dots,\lambda_n)={\rm Id}
+
\mu\sum_{j:j\neq i}r^{(ij)}(\lambda_{ij})
+O(\mu^2).
\ee
Since $\Sigma_1^n\partial_if=0$, we have
\be
\sum_{j:j\neq i}(\partial_i-\partial_j)f=n\partial_if,
\ee
and therefore
\be
T_i(\lambda_1,\dots,\lambda_n)=
{\rm Id}+
\frac n\kappa\mu
\left(
\partial_i+\sum_{j:j\neq i}\frac\kappa n
\sigma_{\lambda_{ij}}(x_{ij})\hat s_{ij}
+\sum_{j:j\neq i}
\frac\kappa n(\rho(\lambda_{ij})-\rho(x_{ij}))\right).
\ee
The claim follows then from the relations
\bea
\partial_ig_m(x)&=&m\sum_{j:j\neq i}\left(\rho(\lambda_{ij})-
\rho(x_{ij})\right),\\
\hat s_{ij}g_m&=&(-1)^mg_m\hat s_{ij},
\eea
and the elementary fact that $m=(-1)^kk$ obeys $(-1)^mm=k$.
\end{proof}
\subsubsection*{6. Quantum $n$-body problems}

Trigonometric and rational Dunkl operators in the $A_{n-1}$
are used
in the theory of integrable quantum $n$-body problems
of the Calogero-Sutherland type, see \cite{He},
\cite{Ch}. Let us consider the
trigonometric Dunkl operators (in a slightly different
normalization)
\be
\nabla_i=\partial_i+\sum_{j:j\neq i}
k(\coth(x_i-x_j)-1)\hat s_{ij}.
\ee
Then the operators $L_j={\rm Res}\sum_{i=1}^n(\nabla_i)^j$,
$j=1,\dots,n$ form a set of pairwise commuting $S_n$-invariant
differential operators, and generate the algebra
of $S_n$-invariant differential operators commuting
with the Schr\"odinger operator
\be
L_2=\sum_{i=1}^n\partial_i^2-k(k+1)\sum_{i\neq j}\frac 1{\sinh^2(x_i-x_j)}.
\ee
Here ${\rm Res}\,M$ is the differential operator whose
restriction to $S_n$-invariant functions coincides with
the differential-difference operator $M$.

In the elliptic case, we are lead to consider the commuting operators
$M_j(\lambda)=\sum_{i=1}^n(\nabla^\lambda_i)^j$, where
$\nabla_i^\lambda$ is the elliptic Dunkl operator \Ref{edo}.
These operators are not $S_n$-invariant. Instead, they obey
$\hat w M_j(\lambda)\hat w^{-1}=M_j(w(\lambda))$. However,
let us consider the singular limit when $\lambda$ tends to
the symmetric point 0,
for $j=2$. We use the shorthand notation $x_{ij}=x_i-x_j$.
\bea
M_2(\lambda)&=&
\sum_{i=1}^n
(\nabla_i^\lambda)^2\\
 &=&\sum_{i=1}^n\partial_i^2
+k\sum_{i\neq j}\sigma'_{\lambda_{ij}}(x_{ij})\hat s_{ij}
+k^2\sum_{i\neq j}
\sigma_{\lambda_{ij}}(x_{ij})\sigma_{\lambda_{ij}}(x_{ji}).
\eea
In this calculation, we used the functional equation
of Prop.\ \ref{aa} (iii) for $\sigma_\lambda$ and the fact that
$\sigma_{-\lambda}(-x)=-\sigma_\lambda(x)$. Since
\be
\sigma_\lambda(x)\sigma_\lambda(-x)=\wp(\lambda)-\wp(x),
\ee
and $\lim_{\lambda\to 0}\sigma'_\lambda(x)=-\wp(x)-2\eta_1$,
(see the Appendix)
we see that
\be M_2(\lambda)-\sum_{i\neq j}\wp(\lambda_{ij})\ee
has a limit
as $\lambda\to 0$:
\be
L_2=\lim_{\lambda\to 0}
(M_2(\lambda)-\sum_{i\neq j}\wp(\lambda_{ij})
=\sum_{i=1}^n\partial_i^2
-k(k+1)\sum_{i\neq j}\wp(x_i-x_j) +{\rm const}.
\ee
This is (minus) the Schr\"odinger operator of the
so-called elliptic Calogero--Moser integrable $n$-body problem
(see, e.g., \cite{OP}).

It is reasonable to conjecture that the
higher $S_n$-invariant differential operators
commuting with $L_2$ can be obtained as $\lambda\to 0$
limits of suitable equivariant polynomials in
$\nabla_i^\lambda$ with $\lambda$ dependent coefficients.

For integer $k$ there is a conjecture of one of the authors
(see \cite{ChVe})
that there are additional integrals of motion, such that
the whole ring of quantum integral is supercomplete.
We hope that elliptic Dunkl operators will help
to prove it.

Another interesting problem is to understand
the analogue of Opdam's shift operator in the elliptic case
(cf.\ \cite{He2}). The one-dimensional case shows that it can not be a
pure differential operator, because the genera of the spectral
curves for Lam\'e operators depend on the integer parameter $k$.

In the $B_2$ case our results imply the quantum integrability
of the system with Hamiltonian
\be
H=-\triangle+F(x)+F(y)+G(x+y)+G(x-y),
\ee
where $F=f'-f^2$, $G=g'-g^2$, and $f$, $g$, are given
by the formula \Ref{vv4}. The commuting operator
$K$ has the form Res$(\nabla_1^2\nabla_2^2)$, where
$\nabla_i$ are the corresponding generalized Dunkl operators
\Ref{e6}.
The quantum integrability of this systems was
independently established in \cite{OOS}.

It would be interesting to understand the relation
between this construction and the recent construction
of elliptic Dunkl operators of \cite{Ch2}, which are
formal infinite linear combinations
 of affine Weyl group reflections.

\paragraph{Acknowledgments.} Two of us (V. B. and A. V.)
are grateful to the University of Maryland at College
Park and especially to Prof.\ S. P. Novikov for the
hospitality during February 1994, when this work was
completed. G. F. is grateful to IHES, where part of
this work was done, for hospitality, and
thanks V. Pasquier for explanations and discussions.

\appendix

\subsubsection*{Appendix: Properties of the function
$\sigma_\lambda(z)$}

We give some properties of the function $\sigma_\lambda(z)$
\Ref{sigma}. This function can be expressed in terms
of the Weierstrass functions (see, e.g., \cite{WW})
with periods $1$, $\tau$:
\be
\sigma_\lambda(z)=
\frac
{\sigma(z-\lambda)}
{\sigma(z)\sigma(-\lambda)}
e^{2\eta_1z\lambda},\qquad \eta_1=\zeta(\frac12).
\ee

\begin{proposition}\label{aa}
\begin{enumerate}
\item[{\rm (i)}] $\sigma_\lambda(z)=-\sigma_{-\lambda}(-z)$,
\item[{\rm (ii)}] $\sigma_\lambda(z)=-\sigma_z(\lambda)$,
\item[{\rm (iii)}] $\sigma_\lambda(z)\sigma_\mu(w)-
\sigma_{\lambda+\mu}(w)\sigma_{\lambda}(z-w)
-\sigma_\mu(w-z)\sigma_{\lambda+\mu}(z)=0$,
\item[{\rm (iv)}] $\sigma_\lambda(z)\sigma_{-\lambda}(z)
=\wp(z)-\wp(\lambda)$,
\item[{\rm (v)}] $\lim_{\lambda\to 0}\sigma'_\lambda(z)
=-\wp(z)-2\eta_1$.
\end{enumerate}
\end{proposition}

\begin{proof}
(i) and (ii) are obvious. The functional equation
(iii) follows from the fact the left hand side is
1-periodic in, say, $z$, and is multiplied by
$\exp(2\pi i\lambda)$ as $z\to z+\tau$. Moreover,
the singularity as $z\to 0$ (and thus as $z\to m+n\tau$,
$n$, $m\in\Z$, by quasiperiodicity) is removable since
$\sigma_\lambda(z)=1/z+O(1)$. Thus we have an entire
1-periodic function of $z$ with non-trivial multipliers
as $z\to z+\tau$. Such a function must vanish.

\noindent (iv) The left hand side is an even elliptic
function with double pole with unit coefficient at the
origin, and no other pole in the period parallelogram.
Thus it is of the form $\wp(z)+\psi(\lambda)$, for
some function $\psi$. Then use $\sigma_\lambda(\lambda)=0$
to determine $\psi$.

\noindent (v)
Recall that $\sigma'(z)=\zeta(z)\sigma(z)$,
$\zeta'(z)=-\wp(z)$ and $\sigma(z)=z+O(z^5)$. We have
\bea
\frac d{dz}
\frac
{\sigma(z-\lambda)}
{\sigma(z)\sigma(-\lambda)}
&=&
\frac{\sigma(z-\lambda)\left(
\zeta(z-\lambda)-\zeta(z)\right)}
{\sigma(z)\sigma(-\lambda)}\\
&=&\zeta'(z)+O(\lambda).
\eea
Thus $\sigma'_\lambda(z)=-\wp(z)+O(\lambda)
+2\eta_1\lambda\sigma_\lambda(z)$,
but $\lambda\sigma_\lambda(z)\to -1$ as $\lambda$ tends to 0.
\end{proof}

\noindent {\em Remark.} Note that the functional
equation (iii) is a special case ($A_2$) of Theorem \ref{tfa}.

\end{document}